\pdfoutput=1
\documentclass{article}
\usepackage{cite}
\usepackage{spconf,amsmath,graphicx}
\usepackage{algorithm}
\usepackage{algorithmic}

\title{Defense against adversarial attacks on \\spoofing countermeasures of ASV}
\name{Haibin Wu$^{1^{*}}$, Songxiang Liu$^{2^{*}}$\thanks{* Equal contribution. Songxiang Liu and Helen Meng were supported by the General Research Fund from the Research Grants Council of Hong Kong SAR Government (Project No. 14208718). Haibin Wu and Hung-yi Lee were supported by the Ministry of Science and Technology of Taiwan.}, Helen Meng$^2$, Hung-yi Lee$^1$}

\address{
  $^1$ Graduate Institute of Communication Engineering, National Taiwan University \\
$^2$Human-Computer Communications Laboratory, 
  The Chinese University of Hong Kong \\ 
  }

\begin{document}
%
\maketitle
\begin{abstract}
Various forefront countermeasure methods for automatic speaker verification (ASV) with considerable performance in anti-spoofing are proposed in the ASVspoof 2019 challenge. However, previous work has shown that countermeasure models are vulnerable to adversarial examples indistinguishable from natural data. A good countermeasure model should not only be robust against spoofing audio, including synthetic, converted, and replayed audios; but counteract deliberately generated examples by malicious adversaries. In this work, we introduce a passive defense method, spatial smoothing, and a proactive defense method, adversarial training, to mitigate the vulnerability of ASV spoofing countermeasure models against adversarial examples. This paper is among the first to use defense methods to improve the robustness of ASV spoofing countermeasure models under adversarial attacks. The experimental results show that these two defense methods positively help spoofing countermeasure models counter adversarial examples.
\end{abstract}
\section{Introduction}
\label{sec:intro}


Automatic speaker verification, abbreviated as ASV, refers to the task of ascertaining whether an utterance was spoken by a specific speaker. ASV is undisputedly a crucial technology for biometric identification, which is broadly applied in real-world applications like banking and home automation. 
Considerable performance improvements in terms of both accuracy and efficiency of ASV systems have been achieved through active research in a diversity of approaches \cite{kanagasundaram2011vector,garcia2011analysis,senior2014improving,reynolds2000speaker,heigold2016end,lei2014novel}. 
\cite{reynolds2000speaker} proposed a method that use the Gaussian mixture model to extract acoustic features and then apply the likelihood ratio for scoring. An end-to-end speaker verification model that directly maps an utterance into a verification score is proposed by \cite{heigold2016end} to improve verification accuracy and make the ASV model compact and efficient. 

Recently, improving the robustness of ASV systems against spoofing audios, such as synthetic, converted, and replayed audios, has attracted increasing attention.
The automatic speaker verification spoofing and countermeasures challenge\cite{wu2015asvspoof,kinnunen2017asvspoof,todisco2019asvspoof}, which is now in its third edition, aims at developing reliable spoofing countermeasures which can counteract the three kinds of spoofing audios mentioned above. The ASVspoof 2019 takes both logical access (LA) and physical access (PA) into account. The LA scenario contains artificially generated spoofing audios by modern text-to-speech and voice conversion models, and the PA scenario contains replayed audios.
A variety of state-of-the-art countermeasure methods that aim at anti-spoofing for ASV models are proposed, and these have achieved considerable performance level for anti-spoofing \cite{gomez2019gated,zeinali2019detecting,lai2019assert,lavrentyeva2019stc,das2019long}. However, whether these countermeasure models can defend against deliberately generated adversarial examples remain to be investigated.

Adversarial examples \cite{szegedy2013intriguing} are generated by maliciously perturbing the original input with a small noise. The perturbations are almost indistinguishable to humans but can cause a well-trained network to classify incorrectly. Using deliberately generated adversarial examples to attack machine learning models is called adversarial attack. Previous work has shown that image classification models are subject to adversarial attacks \cite{szegedy2013intriguing}. The spoofing countermeasure models for ASV learned by the backpropagation algorithm also have such intrinsic blind spots to adversarial examples \cite{Liu2019attackasv}. These intrinsic blind spots must be fixed to ensure safety.

To mitigate the vulnerability of spoofing countermeasure models to adversarial attacks, we introduce a passive defense method, namely spatial smoothing, and a proactive defense method, namely adversarial training. Two countermeasure models in ASVspoof 2019 \cite{zeinali2019detecting,lai2019assert} are constructed, and we implement adversarial training and spatial smoothing to improve the reliability of these two models. This work is among the first to explore defense against adversarial attacks for spoofing countermeasure models.

This paper is organized as follows. In section 2, we introduce the procedure of adversarial example generation. Section 3 gives the detailed structure of two countermeasure models for subsequent experiments. In Section 4, we describe two defense approaches, namely spatial smoothing and adversarial training. The experimental result and analysis are shown in Section 5. 
Finally, conclusion and future work are given in Section 6.

\section{Adversarial Example Generation}

\subsection{Adversarial Example Generation}
We can generate adversarial examples by adding a minimally perceptible perturbation to the input space. The perturbation is found by solving an optimization problem. There are two kinds of adversarial attacks: targeted attacks and nontargeted attacks. Targeted attacks aim at maximizing the probability of a targeted class which is not the correct class. Nontargeted attacks aim at minimizing the probability of the correct class. We focus on targeted attacks in this work. Specifically, to generate adversarial examples, we fix the parameters $\theta$ of a well-trained model and perform gradient descent to update the input. Mathematically, we want to find a sufficiently small perturbation $\delta$ that satisfies (see Equation \ref{eq:optimization}):
\begin{equation}
\label{eq:optimization}
\begin{split}
    \tilde{x} = x + \delta, \\
    f_{\theta}(x) = y, \\
    f_{\theta}(\tilde{x}) = \tilde{y}, \\
    \delta \in \Delta,
\end{split}
\end{equation}
where $f$ is a well-trained neural network parameterized by $\theta$, $x \in R^{N}$ is the input data with dimensionality $N$, $y$ is the true label corresponding to $x$, $\tilde{y}$ is a randomly selected label where $\tilde{y} \not=y$, $\tilde{x} \in R^{N}$ is the perturbed data, $\delta \in R^{N}$ is a small perturbation and $\Delta$ is the feasible set of $\delta$. Finding a suitable $\delta$ is a constrained optimization problem and we can use descent method to solve it. $\Delta$ can be a small $l_{\infty}$-norm ball:
\begin{equation}
\label{eq:l-norm}
\begin{split}
   \Delta = \{\delta| \, ||\delta||_{\infty}\leq \epsilon\},
\end{split}
\end{equation}
where $\epsilon \geq 0$ and $\epsilon \in R$. The constraint in Equation \ref{eq:l-norm} is a box constrain and clipping is used to make the solution $\tilde{x}$ feasible. We choose the feasible set $\Delta$ as shown in Equation \ref{eq:l-norm}.

The projected gradient descent method, abbreviated as PGD method is an iterative method for adversarial attack and has shown effective attack performance in various tasks \cite{madry2017towards}. In this work, the PGD method is introduced to generate adversarial examples. The PGD method is specified in Algorithm~\ref{code:PGD}. In Algorithm~\ref{code:PGD}, $x_{K}$ is the returned adversarial example, the clip() function applies element-wise clipping to make sure $||x_{k} - x||_{\infty}\leq \epsilon$ and $\epsilon \in {R^{+}}$.
\begin{algorithm}[h]
\caption{Projected Gradient Descent Method}
\begin{algorithmic}[1]
\label{code:PGD}
\REQUIRE $x$ and $y$, input and its corresponding label. $\tilde{y}$ is a selected label and $\tilde{y} \not=y$. $\alpha$, step size. $K$, the number of iterations.
\STATE Initialize $x_{0} = x$;
\FOR{$k=0$; $k<K$; $k++$ }
\STATE $\hat{x_{k+1}} = \text{clip}(x_{k} + \alpha \cdot  \text{sign}(\nabla_{x_k} \text{Loss}(\theta, x_{k}, \tilde{y})))$;
\IF{$\text{Loss}(\theta, x_{k+1}, \tilde{y})<\text{Loss}(\theta, x_{k}, \tilde{y})$}
\STATE $x_{k+1} = \hat{x_{k+1}}$
\ELSE
\STATE $x_{k+1} = x_{k}$
\ENDIF
\ENDFOR
\RETURN $x_{K}$;
\end{algorithmic}
\end{algorithm}
\newline


\section{ASV Spoofing Countermeasure Models}
Inspired by the ASV spoofing countermeasure models in the ASVspoof 2019 challenge\cite{todisco2019asvspoof,zeinali2019detecting,lai2019assert}, we construct two kinds of single models to conduct defense methods.
The description of these two models will be given in the subsequent parts.

\subsection{VGG-like Network}
The VGG network, a model made up of convolution layers and pooling layers, has shown remarkable performance in image classification.
\cite{zeinali2019detecting} studied VGG from the perspective of automatic speaker verification and proposed a VGG-like network with good performance on anti-spoofing for ASV. 
Based on this finding, we modified VGG to address anti-spoofing and the modified network structure is shown in Table~\ref{tab:vgg}.
\begin{table}[htb!]
    \centering
    \small
    \vspace{-0.5cm}
    \caption{VGG-like network architecture.}
    \begin{tabular}{c|c|c}
    \hline
   Type &  Filter & Output  \\
    \hline 
    Conv2D-1-1     & $3\times3$    & $64\times600\times257$  \\
    MaxPool-1 & $2\times2$                       & $64\times300\times128$  \\
    \hline
    Conv2D-2-1     & $3\times3$    & $128\times300\times128$  \\
    MaxPool-2 & $2\times2$                       & $128\times150\times64$  \\
    \hline
    Conv2D-3-1     & $3\times3$    & $256\times150\times64$  \\
    Conv2D-3-2     & $3\times3$    & $256\times150\times64$  \\
    MaxPool-3 & $2\times2$                       & $256\times75\times32$  \\
    \hline
    Conv2D-4-1     & $3\times3$    & $512\times75\times32$  \\
    Conv2D-4-2     & $3\times3$    & $512\times75\times32$  \\
    MaxPool-4 & $2\times2$                       & $512\times37\times16$  \\
    \hline
    Conv2D-5-1     & $3\times3$    & $512\times37\times16$  \\
    Conv2D-5-2     & $3\times3$    & $512\times37\times16$  \\
    MaxPool-5 & $2\times2$                       & $512\times18\times8$  \\
    \hline
    Avgpool      & $-$                       & $512\times7\times7$  \\
    Flatten   & $-$                & 25088  \\
    \hline
    FC               & $-$                & 4096\\
    FC               & $-$                & 4096\\
    FC(softmax)       & $-$                & 2\\
    \hline
    \end{tabular}
    \label{tab:vgg}
\end{table}
\subsection{Squeeze-Excitation ResNet model}
Lai et al. \cite{lai2019assert} proposed the Squeeze-Excitation ResNet model (SENet) to address anti-spoofing for ASV. The system proposed by \cite{lai2019assert} ranked 3rd and 14th for the PA and LA scenarios respectively in the ASVspoof 2019 challenge. However, \cite{Liu2019attackasv} successfully attacked the SENet by deliberately generated adversarial examples. Hence, this work seeks to improve the robustness of SENet with two defense methods elaborated below.

\section{Defense Methods}
There are two kinds of defense methods against adversarial attacks: passive defense and proactive defense. Passive defense methods aim at countering adversarial attacks without modifying the model. Proactive defense methods train new models which are robust to adversarial examples. Two defense methods are introduced in this section: spatial smoothing which is inexpensive and complementary to other defense methods and adversarial training.
\subsection{Spatial Smoothing}
Spatial smoothing (referred as "filtering") has been widely used for noise reduction in image processing. It is a method that uses the nearby pixels to smooth the central pixel. There are a variety of smoothing methods based on different weighting mechanisms of nearby pixels, e.g., median filter, mean filter, Gaussian filter, etc. Take the mean filter as an example, a slicing window moves over the picture and the central pixel in the window will be substituted by the mean of the values within the slicing window. 

 Spatial smoothing was introduced by \cite{xu2017feature} to harden image classification models by detecting malicious generated adversarial examples. Implementing smoothing does not need extra training effort, so we use this inexpensive strategy to improve the robustness of well-trained ASV models.

\begin{algorithm}[h]
\caption{}
\begin{algorithmic}[1]
\label{code:adversarial training}
\REQUIRE $X$ and $Y$, set of paired audio and its corresponding labels. $\theta$, network parameters. $T_{1}$, normal training epoch. $T_{2}$, adversarial training epoch. $N$, number of training examples, $b$, batch size.

\STATE Initialize $\theta$.
\FOR{$t=0$; $t<T_{1}$; $t++$ }
\FOR{$i=0$;$i<N/b$;$i++$ }
\STATE Get $\{(x_{i},y_{i})\}_{i=1}^{i=b}$ from $\{X,Y\}$;
\STATE Update $\theta$ using gradient decent with respect to $\{(x_{i},y_{i})\}_{i=1}^{i=b}$;
\ENDFOR
\ENDFOR
\WHILE{\{$t<=T_{2}$ \& $\theta$ not converged\}}
\FOR{$i=0$; $i<N/b$; $i++$}
\STATE Get $\{(x_{i},y_{i})\}_{i=1}^{i=b}$ from $\{X,Y\}$;
\STATE Generate adversarial examples $\{(\tilde{x_{i}})\}_{i=1}^{i=b}$ by PGD method;
\STATE Update $\theta$ using gradient decent with respect to $\{(\tilde{x_{i}},y_{i})\}_{i=1}^{i=b}$;
\ENDFOR
\ENDWHILE
\RETURN $\theta$;
\end{algorithmic}
\end{algorithm}

\subsection{Adversarial Training}
Adversarial training, which utilizes adversarial examples and injects them into training data, was introduced in \cite{goodfellow2014explaining} to mitigate the vulnerability of deep neural networks against adversarial examples. 
Adversarial training can be seen as a combination of an inner optimization problem and an outer optimization problem where the goal of the inner optimization is to find imperceptible adversarial examples and the goal of outer optimization is to fix the blind spots.
In this work, we also employ adversarial training. First, we use clean examples to pre-train the countermeasure models for $T_{1}$ epochs. Then we do adversarial training for $T_{2}$ epochs. The detailed implementation procedure is shown in Algorithm~\ref{code:adversarial training}.

\section{Experiment}
\subsection{Experiment Setup}
In this paper, we use the LA partition of the ASVspoof 2019 dataset \cite{todisco2019asvspoof}. The LA partition is divided into training, development and evaluation sets. The training and development sets are generated by the same kinds of TTS or VC models while the evaluation set contains examples generated by different kinds of TTS or VC models. We trained and then tested on the development set to ensure similar distributions between the datasets. Raw log power magnitude spectrum computed from raw audio waveform is used as acoustic features. A Hamming window of size 1724 and step-size of 0.001s is used to extract FFT spectrum. We use only the first 600 frames of each utterance for training and testing. We do not employ additional preprocessing methods such as dereverberation or pre-emphasis.

The network structures of the two countermeasure models were as described in Section 3. During the experiment, we first use the training data to pre-train the countermeasure models. Then the PGD method as shown in Algorithm~\ref{code:PGD} is adopted to generate adversarial examples for the well-trained countermeasure models. When we run the PGD method, $\epsilon$ is set to 5, $K$ is set to $10$ and $\alpha$ is set to $0.5$. Then we measure the performance of well-trained countermeasure models by the generated adversarial examples with and without filters. Three kinds of filters including median filter, mean filter and Gaussian filter are implemented. Then we use adversarial training to train the countermeasure model for $T_{2}$ epochs as shown in Algorithm~\ref{code:adversarial training}. After adversarial training, we evaluate the testing accuracy of countermeasure models for adversarial examples.

\subsection{Results and Analyses}
\subsubsection{Spatial Smoothing}
After we pre-train VGG and SENet for $T_{1}$ epochs, we evaluate the testing accuracy of these two models.
According to Table~\ref{tab:accuracy before}, both SENet and VGG achieve high testing accuracy in the testing data which is not perturbed. However, when we test the two models with adversarial examples, the testing accuracy drops drastically. When we apply spatial smoothing to the adversarial examples and then evaluate the performance, the adversarial attack becomes ineffective as there is a great increase in testing accuracy. All three kinds of spatial filters have considerable performance in improving the robustness of countermeasure models against adversarial examples. The improvement obtained with Gaussian filters is much less than the other two filters. 

We attempt to explain the contribution of spatial smoothing contributes to spoofing countermeasure model to be robust against adversarial examples. In the adversarial attack scenario, an adversary has full access to a well-trained model but can not alter the parameters of the model. Now, assuming that the adversary is not aware of the existence of spatial smoothing which will be implemented to the input data before the input is thrown into the model. The adversary attempts to find an imperceptible noise which will cause the well-trained model to classify incorrectly by the PGD method and add it to the input. However, 
the deliberately generated perturbation will be countered by spatial smoothing and the adversarial attack becomes invalid. 

\newcommand{\tabincell}[2]{\begin{tabular}{@{}#1@{}}#2\end{tabular}}  
\begin{table}[htb!]
    \centering
    \small
    \vspace{-0.5cm}
    \caption{Testing accuracy of VGG and SENet before adversarial training. }
    \begin{tabular}{c|p{1.2cm}|p{1.2cm}}
    \hline
    & SENet & VGG \\
    \hline
    Normal examples & 99.97\% & 99.99\%\\
    \hline
    Adversarial examples & 48.32\% & 37.06\%\\
    \hline
    \tabincell{c}{Adversarial examples \\ + median filter}& 82.00\% & 92.72\% \\
    \hline
    \tabincell{c}{Adversarial examples \\ + mean filter}& 82.39\% & 93.95\% \\
    \hline
    \tabincell{c}{Adversarial examples \\ + Gaussian filter}& 78.93\% & 84.39\%\\
    \hline 
    \end{tabular}
    \label{tab:accuracy before}
\end{table}

\subsubsection{Adversarial Training}
As shown in Table~\ref{tab:accuracy after}, the testing accuracy for adversarial examples of SENet increases from 48.32\% to 92.40\% while the testing accuracy for normal examples changes little after adversarial training.
We can see a similar phenomenon for VGG. According to Table~\ref{tab:accuracy after}, adversarial training does improve the robustness of VGG and SENet.

\begin{table}[htb!]
    \centering
    \small
    \vspace{-0.5cm}
    \caption{Testing accuracy of VGG and SENet after adversarial training.}
    \begin{tabular}{c|p{1.2cm}|p{1.2cm}}
    \hline
    & SENet & VGG \\
    \hline
    Normal examples & 99.75\% & 99.99\%\\
    \hline
    Adversarial examples & 92.40\% & 98.60\%\\
    \hline
    \tabincell{c}{Adversarial examples \\ + median filter}& 93.74\% & 98.96\% \\
    \hline
    \tabincell{c}{Adversarial examples \\ + mean filter}& 93.76\% & 99.24\% \\
    \hline
    \tabincell{c}{Adversarial examples \\ + Gaussian filter}& 83.72\% & 87.22\%\\
    \hline
    \end{tabular}
    \label{tab:accuracy after}
\end{table}

Traditional supervised training does not address the chosen models to be robust to adversarial examples.
So the well-trained models by traditional supervised learning may be sensitive to changes in its input space and thus have vulnerable blind spots that can be attacked by a malicious adversary.
During the training stage, adversarial attacks should be taken into account by training on a mixture of data which contains not only clean examples but also adversarial examples to regularize and make the model insensitive on all data points within the $\epsilon$ max norm box.
After doing that, it is hard for malicious adversaries to generate adversarial examples to attack the model.
Adversarial training largely samples adversarial examples within the $\epsilon$ max norm box to augment the training set.
The results in Table~\ref{tab:accuracy after} illustrate that it is feasible and practical to train a robust countermeasure model using adversarial training. 
 

\subsubsection{Adversarial Training + Spatial Smoothing}
We combine spatial smoothing and adversarial training and the experiment results are shown in Table~\ref{tab:accuracy after}. We observe that equipping adversarial training with median filters or mean filters increase the testing accuracy for adversarial examples, as compared to solely using adversarial training. But adding Gaussian filters decreases the testing accuracy. Hence, Median filters and mean filters are more desirable filters than Gaussian filters in our experiment setting.


\section{Conclusion}
In this paper, two kinds of defense methods, namely spatial smoothing and adversarial training, are introduced to improve the robustness of spoofing countermeasure models under adversarial attacks. We implement two countermeasure models, i.e., VGG and SENet and augment them with defense methods. The experiment results show both spatial smoothing and adversarial training enhance robustness of the models against adversarial attacks.

For future work, we will introduce powerful defense methods, such as ensemble adversarial training \cite{tramer2017ensemble}, to make spoofing countermeasure models more robust to adversarial audios generated from testing data having different distribution with training data.

\bibliographystyle{IEEEbib}
\bibliography{strings,refs}

\end{document}